\documentclass[twocolumn,amsmath,amssymb,floatfix]{revtex4}

\usepackage{graphicx}
\usepackage{dcolumn}
\usepackage{bm}

\graphicspath{{Figures/}}
\setkeys{Gin}{width=\linewidth}

\begin{document}

\title{Strong spin--orbit interactions and weak antilocalization\\ in carbon doped p-type GaAs heterostructures}

\author{Boris Grbi\'{c}$^{*}$, Renaud Leturcq$^{*}$, Thomas 
Ihn$^
{*}$,
Klaus Ensslin$^{*}$, Dirk Reuter $^{+}$, and Andreas D.
Wieck$^{+}$}

\affiliation{$^{*}$Solid State Physics Laboratory, ETH Zurich,
  8093 Zurich, Switzerland,\\$^{+}$Angewandte Festk\"orperphysik,
Ruhr-Universit\"{a}t Bochum, 44780 Bochum, Germany}

\begin{abstract}

We present a comprehensive study of the low-field magnetoresistance in carbon 
doped p-type GaAs/AlGaAs heterostructures aiming at the investigation of spin--orbit interaction effects.
The following signatures of
exceptionally strong spin-orbit interactions are simultaneously
observed: a beating in the Shubnikov-de Haas oscillations, a
classical positive magnetoresistance due to the presence of the
two spin-split subbands, and a weak anti-localization dip in the
magnetoresistance. The spin-orbit induced splitting of the heavy 
hole subband at the Fermi level is determined to be around 30 \% of
the total Fermi energy. The phase coherence length of holes of around 
2.5\,$\mu$m at a temperature of 70\,mK, extracted from weak
anti-localization measurements, is promissing for the fabrication of
phase-coherent p-type nanodevices.

\end{abstract}

\maketitle

 \section{Introduction}

Two-dimensional (2D) systems with strong spin-orbit 
interactions
(SOI) are promising for the realization of spintronics devices
due to the fact that in such systems the electron (hole) spin
could be affected, not only by magnetic, but also by electric
fields \cite{Dressel55,Bychkov84}. SOI are expected to be very strong in p-type GaAs
heterostructures, due to the
high effective mass of holes
\cite{Winkler03} which makes the ratio of the SOI energy and the kinetic energy larger than in the conduction band. As a result of the SOI, the heavy hole subband
in GaAs is split into two subbands even in the 
absence of an external magnetic field.

In magnetotransport experiments the existence of two
spin-split subbands with different mobilities results in a classical positive magnetoresistance. In addition, a beating can be observed in Shubnikov-de 
Haas (SdH) oscillations, because the Landau levels of the two non-equally
populated subbands give rise to magnetoresistance oscillations
with slightly different $1/B$-periodicities \cite{Zawadzki04}. While these two signatures can be observed in any two-subband system, the spin
splitting due to SOI can be unambiguously identified and characterized by measurements of the weak anti-localization effect.

Weak localization is a quantum mechanical effect which arises
from the constructive interference between time reversed partial
waves of the charge carriers in disordered materials. It leads to an
enhanced probability of carrier backscattering and therefore to an
enhanced longitudinal resistivity. This interference effect is relevant for diffusive orbits up
to the length scale $l_\varphi$, the phase-coherence length. The application of a
magnetic field normal to the plane of carrier motion breaks the time
reversal symmetry, suppresses the weak localization, and therefore leads to
a negative magnetoresistance at low magnetic fields around $B=0$ \cite{Beenakker}.

In systems with strong SOI the spin dynamics of the carriers is
coupled to their orbital motion and the interference of time-reversed paths
has consequences beyond the weak localization effect. As the
spin experiences a sequence of scattering events along its
path, the spin orientation is randomized on a characteristic length scale $l_\mathrm{so}$. The stronger the SOI, the smaller is $l_\mathrm{so}$. At $B=0$, the interference of time reversed paths leads to a reduction of the backscattering probability below its classical value \cite{Bergmann82}, an effect called weak anti-localization, if $l_\mathrm{so}\ll l_\varphi$ (strong SOI).
It manifests itself as a positive (rather than a negative) magnetoresistance at small fields around $B=0$ \cite{Hikami}.

Weak anti-localization was experimentally observed by
Bergmann in thin metallic films \cite{Bergmann82b}. As 
the strength of SOI is increased, a transition from
weak localization to weak anti-localization is observed. Weak
anti-localization was subsequently observed also in 
semiconductor heterostructures \cite{Poole82, Dresselhaus92}. A smaller
zero-field anti-localization resistance minimum superimposed on a larger weak 
localization
peak was seen in the magnetoresistance of an inversion layer of
InP \cite{Poole82}, and an n-type GaAs/AlGaAs heterostructure
\cite{Dresselhaus92}. A fully developed anti-localization minimum 
was
observed by Chen {\em et al.} in the magnetoresistance of an InAs
quantum well \cite{Chen93}. Koga {\em et al.} demonstrated the
transition from a zero-field weak localization maximum to a weak
anti-localization minimum by tuning the symmetry of an InGaAs
quantum well (QW) wih a metallic top-gate \cite{Koga02}.

Weak anti-localization is expected to be particularly expressed
in the case of p-type GaAs heterostructures due to the strong SOI in
these systems. Experimental studies of weak anti-localization 
in Be-doped (100) p-type GaAs heterostructures are reported in
Refs.\,\onlinecite{Pedersen99} and \onlinecite{Yaish01}, and a detailed study of the low-field
magnetoresistance in Si-doped (311) p-type GaAs 
heterostructures
is presented in Ref.\,\onlinecite{Papadakis02}.

Here we report measurements of the
classical magnetoresistance, SdH oscillations,
and weak anti-localization in C-doped p-type GaAs
heterostructures. Weak anti-localization is
typically more pronounced in diffusive, low-mobility samples,
while for the observation of beating in SdH oscillations higher mobility
samples are required. The fact that our sample is in the regime
of intermediate mobilities enables us to simultaneously observe 
both effects and to perform a complementary analysis of spin--orbit interactions in the system.
The observation of a fully developed
anti-localization minimum around $B=0$ clearly demonstrates the
presence of very strong SOI. A phase-coherence time of the holes of
around 190 ps, corresponding to a phase-coherence length of 
2.5\,$\mu$m is extracted from these measurements. 
We investigate the temperature dependence of the phase-coherence
time of holes and find that it obeys a $1/T$
dependence with reasonable accuracy.
Limitations in extracting the spin-orbit scattering time are due to 
the fact that SOI is very strong. It cannot be treated as a
weak perturbation only, as discussed below. 

\section{Sample and measurement setup}

We have studied the low-field magnetoresistance in two
C-doped p-type GaAs heterostructures with the two-dimensional hole gas (2DHG) buried
45\,nm and 100\,nm below the surface. The results obtained from both samples 
are qualitatively the same.
For the sake of clarity, we present here
results obtained on the sample with the 2DHG formed at the interface 100\,nm below the sample surface. The heterostructure
consists of a 5\,nm C-doped GaAs cap layer, followed by a 65\,nm
thick, homogeneously C-doped layer of Al$_{0.31}$Ga$_{0.69}$As
which is separated from the 2DHG by a 30 nm thick, undoped
Al$_{0.31}$Ga$_{0.69}$As spacer layer \cite{Wieck00}. A rectangular Hall bar was
fabricated by standard photolithography. Its width is 100\,$\mu$m and the separation between adjacent voltage leads is 500\,$\mu$m.  Ohmic contacts were formed by evaporating Au and Zn and subsequent annealing at 480$^{o}$C for 2 min.
Afterwards, a homogeneous Ti/Au topgate was evaporated, which
allows to tune the density in the range of $2-3\times 10^{11}$\,cm$^{-2}$.
The average mobility in the sample at $T=70$mK is
160'000 cm$^2$/Vs at a density 3$\times$10$^{11}$ cm$^{-2}$.
The high quality of the investigated sample has been demonstrated by the
observation of the fractional and integer quantum Hall effects, 
as well as by measurements of highly resolved SdH oscillations \cite{Grbic04}.

The Hall bar is fabricated along one of the two main
crystallographic directions in the (100) plane. Measurements at $T=4.2$\,K
on another sample fabricated from the same wafer patterned into an L-shaped Hall
bar have shown that the mobility anisotropy between the two main
crystallographic directions in the (100) plane is less than 25 \%
\cite{GrbicThesis07}, which is significantly less than in
Si-doped (311) p-type GaAs heterostructures \cite{Heremans94}.
Therefore, in contrast to p-type GaAs samples on (311) substrates, where the mobility anisotropy had to be invoked for the interpretation of the low field magnetoresistance  \cite{Papadakis02}, in our measurements on (100) p-type GaAs samples the mobility anisotropy could be neglected.

We have performed four-terminal measurements of the resistivity
using standard low-frequency lock-in techniques. A current of 20\,nA was driven through 
the Hall bar at a frequency of 31\,Hz, and the voltage was measured 
with an integration time of 300 ms. In order to improve the signal-to-noise ratio we
used a voltage amplifier with an amplification of 1000 directly
at the outputs of the cryostat. In order to increase the
signal-to-noise ratio further in measurements of the weak-antilocalization effect,
each data point is the average of 25 samples taken with a temporal separation of 1.5 s.
In this way we reached a noise level of less than 0.03\,$\Omega$ for measured resistances above 200 $\Omega$. For these measurements a special,
home-built power supply for the magnet was used, and the magnetic
field is stepped with a resolution of 40\,$\mu$T.

\section{Beating of Shubnikov-de Haas oscillations}

We have mentioned before that the two spin-split heavy hole subbands
arising as a result of SOI lead to a beating of SdH oscillations.
Figure 1(a) displays a magnetoresistance trace taken in the magnetic field range between -0.1\,T and 1.6\,T showing SdH oscillations. The Fourier analysis of the magnetoresistivity $\rho_{xx}$ vs. $1/B$ [see inset of Fig.\, 1(a) and discussion below] is used to deduce the densities  $N_{1,2}$ of the two spin-split subbands. They are related to the two frequencies $f_{1,2}$ obtained from the Fourier transform of the SdH 
oscillations via $N_{1,2}=(e/h)\cdot f_{1,2}$ \cite {Winkler03, Zawadzki04,
Lu98}.

Three magnetic field regimes can be identified in the raw data, where SdH oscillations exhibit a different behavior. For very low fields in the interval $0.2\,\mbox{T}<B<0.4\,\mbox{T}$, only SdH oscillations originating from the higher mobility spin-split 
subband are observed allowing to extract its density. As the
magnetic field is further increased into the region between $0.4\,\mbox{T}<B<2\,\mbox{T}$, the contribution from the second spin-split subband becomes visible in the oscillations. The Fourier transform analysis data in this range results in a spectrum with three peaks corresponding to the populations of each of the two spin-split subbands, and to the 
total density. An example of such a Fourier transform spectrum obtained from data in the range $0.4\,\mathrm{T}<B<1.5\,\mathrm{T}$ is shown in the inset of Fig.\,1(a). Three peaks (at 4.45\,T, 7.8\,T, and 12.3\,T) can be seen. The relation $f_1+f_2=f_{tot}$, which reflects the fact that the two subband densities sum up to the total density, is reasonably satisfied. For magnetic fields above approximately 2\,T (not shown) we observe magnetoresistance oscillations related to the total density in the
system, and only the total density peak is present in the Fourier spectrum.

\begin{figure}
  \begin{center}
    \includegraphics{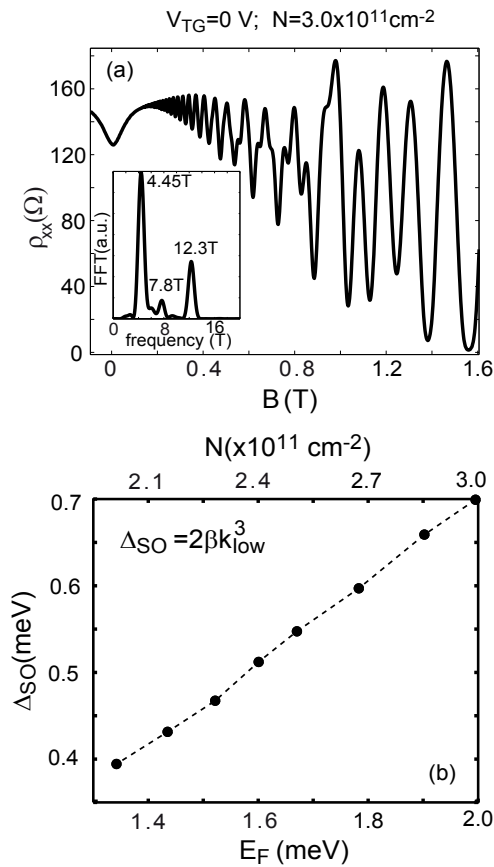}
    \caption{(a) Shubnikov-de Haas oscillations in the
magnetoresistance, with a top gate set to $V_{TG}=0$V, and the
total density $3.0\times 10^{11}$ cm$^{-2}$;
Inset: Fourier transform of the shown magnetoresistance, taken 
in
the B-field range (0.4T, 1.5T) displaying three peaks. (b)
Spin-orbit splitting energy of the heavy hole subband at the
Fermi
level as a function of the Fermi energy in the system and the
total density. }
    \label{fig1NEW}
  \end{center}
\end{figure}

From the three peaks in the Fourier transform shown in the inset of Fig. 1(a) we read the densities of the two spin-split subbands $N_1= 1\times10^{11}$\,cm$^{-2}$, $N_2=1.9\times 10^{11}$\,cm$^{-2}$, and the total density $N=3\times 10^{11}$\,cm$^{-2}$. This corresponds to a relative charge imbalance between the two spin-split subbands $\Delta N/N =0.30$. The strength of the spin-orbit interactions can be quantified using this relative charge imbalance, if a cubic wave vector $k$ dependence $\Delta_\mathrm{SO}=2\beta k_\parallel^3$ is assumed for heavy holes in the (100) plane \cite{Gerchikov92,Winkler03}.
 The two subband's Fermi wavevectors $k_1$ and $k_2$ are different. This
difference increases with increasing spin-orbit interaction. From the general relation $k_i=\sqrt{4\pi N_i}$ we find $k_1= 1.1\times 10^{8}$\,m$^{-1}$, and $k_2=1.6\times 10^{8}$\,m$^{-1}$ at the total density
$N=3\times 10^{11}$\,cm$^{-2}$. The energy splitting of the two
spin-split subbands depends on the $k$-vector where this
splitting is calculated. The values of the spin-splitting energy
which we quote further in the text, are all obtained using the
smaller of the two wavevectors, and therefore represent the 
lower bound for the spin-splitting of the heavy hole subband.

Using the masses of the carriers in the two spin-split subbands determined experimentally in Ref.\,\onlinecite{Grbic04} and the two subband densities, we calculate the spin-orbit coupling parameter from eq.\,(6.39) in Ref.\,\onlinecite{Winkler03} to be
$\beta=2.5\times10^{-28}$ eVm$^3$.
This gives the spin-orbit induced splitting of the heavy hole subband $\Delta_\mathrm{SO}\approx0.7$\,meV at a density $N=3\times 10^{11}$\,cm$^{-2}$. The Fermi
energy for this density is $E_\mathrm{F}=2.0$ meV. As a consequence, the relative strength of the spin-orbit interaction and the kinetic energy is
$\Delta_\mathrm{SO}/E_\mathrm{F} \approx 35\%$. In the gate voltages shown in Fig.\,1(b) the parameter $\beta$ increases with increasing density by about 20\%.

The evolution of the spin-splitting energy $\Delta_\mathrm{SO}$ upon
changing the total density in the system with the metallic top-
gate is shown in Fig. 1(b). It can be seen that for densities in the range $2-3\times 10^{11}$\,cm$^{-2}$ the
spin-splitting energy is in the range of $0.4-0.7$\,meV. Thus, the
relative strength of spin-orbit interactions compared to the
Fermi energy, $\Delta_\mathrm{SO}/E_\mathrm{F}$, is quite large, increasing from 
0.29 to 0.35 with the Fermi energy increasing from 1.35 to 2\,meV. This documents the presence of exceptionally strong SOI in C-doped p-GaAs heterostructure.

\section{Classical positive magnetoresistance}

The longitudinal magnetoresistance of a system with two types of charge carriers with different mobilities is parabolic around zero magnetic field, whereas the Hall resistivity,
contains a small cubic correction at low fields in addition to the usual term linear in $B$. This is a purely classical effect and follows from the standard Drude theory of conductivity \cite{Ashcroft}. If inter-subband scattering between the two subbands is significant, a more complex theory based on the Boltzmann transport equation has to be considered \cite{Zaremba92}. However, the qualitative behavior of the low-field magnetoresistance remains very similar to that obtained using the simpler model neglecting intersubband scattering.

In the transport theory of two-subband systems developed by Zaremba \cite{Zaremba92}, where inter-subband scattering is included, the longitudinal and transverse magnetoresistivity are given by
\begin{eqnarray}
 \rho_{xx}=\frac{m^*}{e^2} \cdot \mathrm{Re}(\frac{1}{\mathrm
{Tr}
\mathbf{N} (\mathbf{K}-i\omega _c \mathbf{I})^{-1}}),\label{rxx}
\\ \rho_{xy}=\frac{m^*}{e^2} \cdot \mathrm{Im}(\frac{1}{\mathrm
{Tr}
\mathbf{N} (\mathbf{K}-i\omega _c \mathbf{I})^{-1}}),\label{rxy}
\end{eqnarray}
where Tr stands for the trace operation, $\mathbf{I}$ is the
$2\times2$ unit matrix, $\mathbf{N}$ is a matrix defined as
$N_{ij}=\sqrt{N_{i}N_{j}}$ ($N_1, N_2$ are the densities of the
two subbands), and $\mathbf{K}$ is the scattering matrix 
\begin{equation*}
\left(\begin{array}{rr} K_1 & -K_{12} \\ -K_{12} & K_2
\end{array} \right),
\end{equation*}
where $K_1, K_2$ are rates quantifying intra-subband scattering, while $K_{12}$ is the inter-subband scattering rate.

Previously, a strong positive magnetoresistance was observed in p-type (311) GaAs heterostructures \cite{Papadakis02}. However, in that case the low-field magnetoresistance could not be fitted satisfactorily  with the two-subband theory, even when intersubband scattering was taken into account. This finding was
attributed to the strong mobility anysotropy in (311) samples.

\begin{figure}
  \begin{center}
    \includegraphics{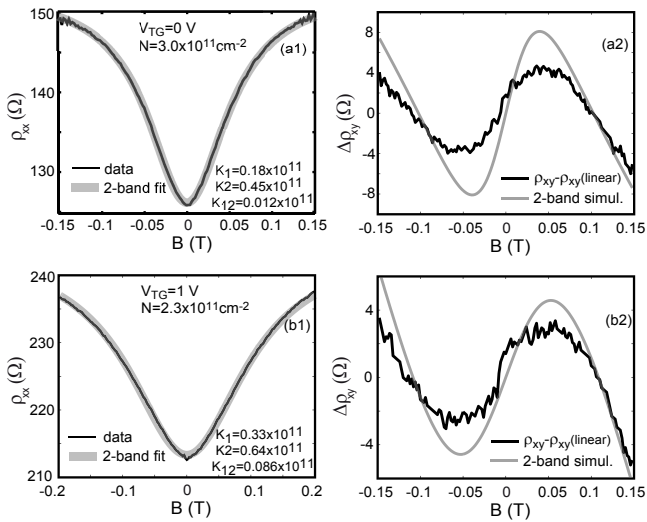}
    \caption{Left column: Fit of the low-field magnetoresistivity with the two-band theory\cite{Zaremba92} (black lines represent the measurement, and thicker gray lines are fitted lines) in the following gate configurations: (a1) $V_\mathrm{TG}=0$, $N= 3.0\times 10^{11}$\,cm$^{-2}$, (b1) $V_\mathrm{TG}=1$\,V, $N= 2.3\times 10^{11}$\,cm$^{-2}$. Right column: Nonlinearity in the low-field Hall
resistivity (black lines are measured data, gray lines are calculated curves,
see text for detailed explanations) in the following gate
configurations: (a2) $V_\mathrm{TG}=0$, $N= 3.0\times 10^{11}$\,cm$^{-2}$, (b2)
$V_\mathrm{TG}=1$\,V, $N= 2.3\times 10^{11}$\,cm$^{-2}$.}
    \label{fig2NEW}
  \end{center}
\end{figure}
Figure 2 shows a strong positive magnetoresistance around $B=0$ in two gate configurations: (a1) $V_\mathrm{TG}=0$, $N=
3.0\times 10^{11}$\,cm$^{-2}$, and (b1) $V_\mathrm{TG}=1$\,V, $N=2.3\times 10^{11}$\,cm$^{-2}$. The black lines correspond to measured data, while the thicker gray lines show the fits following eq.\,(\ref{rxx}) in the range $|B|<0.15\,\mathrm{T}$ for $V_\mathrm{TG}=0$, and $|B|<0.2$\,T for $V_\mathrm{TG}=1$\,V,where SdH oscillations are not yet developed. In the fitting procedure, the densities of the two subbands $N_1, N_2$ are fixed parameters given from the Fourier analysis of the SdH oscillations, whereas the scattering rates $K_1, K_2, K_{12}$ are fitting parameters.

In the configuration $V_\mathrm{TG}=0$, $N= 3.0\times 10^{11}$\,cm$^{-2}$
[Fig. 2(a1)] the scattering rates are $K_1=0.018$\,ps$^{-1}$, $K_2=0.045$\,ps$^{-1}$, and $K_{12}=0.0012$\,ps$^{-1}$. The inter-subband scattering rate is much smaller than the intra-subband scattering rates. The corresponding subband mobilities are $\mu_1=270'000$\,cm$^2$/Vs and
$\mu_2=110'000$\,cm$^2$/Vs. These values explain why in SdH measurements
oscillations arising from the subband with population $N_1$ are observed at lower magnetic fields than those from the subband with
population $N_2$. In the second configuration with $V_\mathrm{TG}=1$\,V, $N=2.3\times 10^{11}$\,cm$^{-2}$ [Fig. 2(b1)] the scattering rates are $K_1=0.033$\,ps$^{-1}$, $K_2=0.064$\,ps$^{-1}$, and $K_{12}=0.0086$\,ps$^{-1}$. Again the
inter-subband scattering rate is about one order of magnitude smaller than
the scattering rates of individual subbands. However, as the density is reduced, we observe that the scattering rates of the individual subbands increase by less than a factor of 2, while the intersubband scattering rate increases by a factor of 7. Such a behavior can be related to the fact that the energy separation $\Delta_\mathrm{SO}$ between the two spin-split bands decreases with density and therefore it is easier for the
carriers to scatter from one subband to the other. By reducing the density, the parabolic feature in the magnetoresistance around $B=0$ becomes broader and shallower.

We have also observed that an increase of the temperature causes a broadening of the 
magnetoresistance minimum around $B=0$, and also an increase of the intersubband 
scattering rate. The intersubband scattering rate increases faster with increasing temperature than the intrasubband scattering rates. This indicates that the presence of the two spin-split subbands in p-type samples might be relevant for the strong temperature dependence of the resistivity even at mK temperatures.

Beside the longitudinal magnetoresistance minimum around $B=0$, the presence of the two spin-split subbands also modifies the Hall resistivity around $B=0$ and introduces non-linear corrections (see eq.\,\ref{rxy}) \cite{Zaremba92}. We have calculated the Hall resistivity using the scattering rates $K_{1}$, $K_{2}$ and $K_{12}$ obtained from the $\rho_{xx}$-fits as input parameters. In order to make these small non-linear corrections to the Hall resistivity visible, we subtract the linear contributions from both the measured data and the calculated $\rho_{xy}$. The result for the measured data (black lines) and the calculated $\rho_{xy}$ (gray lines) is presented in Figs.\,2(a2) (configuration $V_\mathrm{TG}=0$, $N=3.0\times 10^{11}$\,cm$^{-2}$) and 2(b2) (configuration $V_\mathrm{TG}=1$\,V, $N=2.3\times 10^{11}$\,cm$^{-2}$). We find reasonable agreement
between the data and the simulated non-linear corrections of the Hall resistivity.

\section{Weak anti-localization measurements}

Weak (anti)localization effects are observable in lower mobility samples in the diffusive transport regime, if the carrier mean free path is much smaller than the phase-coherence length. In higher mobility samples where $k_\mathrm{F} l_\mathrm{m} \gg 1$ ($k_\mathrm{F}$--Fermi wavevector, $l_\mathrm{m}$--mean free path) localization effects are weaker and harder to resolve. The measured density and mobility values in the investigated sample give $k_\mathrm{F} l_\mathrm{m} \sim 100-200$. Therefore, the magnitude of the localization effects is expected to be very small. In order to
resolve a weak anti-localization peak in the magnetoresistivity we had to determine both the resistance and the magnetic field in the narrow $B$-field range around $B=0$ with very high accuracy.

\begin{figure}
  \begin{center}
    \includegraphics{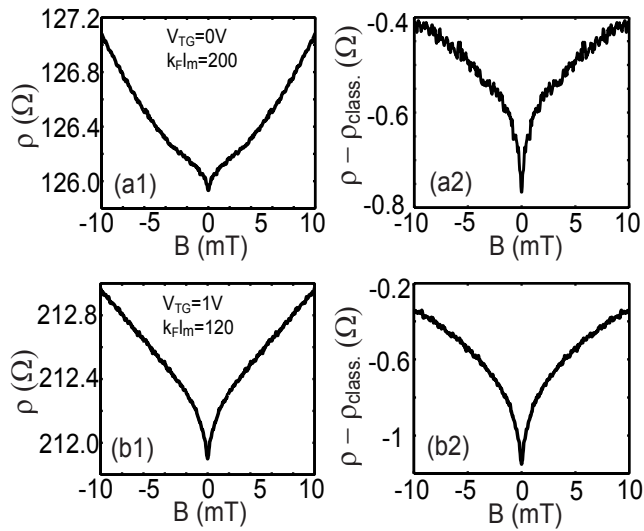}
    \caption{Left column: Raw magnetoresistivity data at $T=65$\,mK in the following gate configurations: (a1) $V_\mathrm{TG}=0$, $N=3\times 10^{11}$\,cm$^{-2}$, $\mu=160'000$\,cm$^{2}$/Vs, (b1) $V_\mathrm{TG}=1$\,V, $N=2.3\times 10^{11}$\,cm$^{-2}$, $\mu=130'000$\,cm$^{2}$/Vs. Right column: Quantum correction of the resistivity obtained after subtraction of the classical two-band positive magnetoresistivity for (a2) $V_\mathrm{TG}=0$, (b2) $V_\mathrm{TG}=1$\,V.}
    \label{fig3NEW}
  \end{center}
\end{figure}
In the left column of Fig.\,3, the raw magnetoresistivity data are presented for the gate-configurations $V_\mathrm{TG}=0$, $N=3\times 10^{11}$\,cm$^{-2}$, $\mu=160'000$\,cm$^{2}$/Vs, $k_\mathrm{F} l_\mathrm{m}=200$ [Fig.\,3 (a1)], and $V_\mathrm{TG}=1$\,V, $N=2.3\times 10^{11}$\,cm$^{-2}$, $\mu=130'000$\,cm$^{2}$/Vs, $k_\mathrm{F} l_\mathrm{m}=120$ [Fig.\,3 (b1)]. In both cases we observe a sharp anti-localization resistance minimum around $B=0$ with a magnitude much smaller than $1\,\Omega$. It can be seen that the magnitude and the width of the anti-localization minimum become larger as the factor $k_\mathrm{F} l_\mathrm{m}$ decreases due to a reduction of the sample mobility and density at positive top gate.

As discussed before a classical magnetoresistance minimum is present around $B=0$ due to the presence of the two spin-split subbands. In order to separate the quantum correction from the low-field magnetoresistivity, we subtract the classical 
positive magnetoresistivity $\rho_\mathrm{class}$ [thick gray lines in Fig.\,2(a1)
and 2(b1)] from the total resistivity $\rho$. The quantum corrections to the resistivity, $\rho-\rho_\mathrm{class}$, are plotted in the right column of Fig.\,3 for both gate configurations.

It can be seen in Fig.\,3(a2,b2) that in both cases a well developed weak anti-localization minimum is present in the low-field magnetoresistance. The fact that the narrow weak anti-localization minimum is not superimposed on a wider weak localization peak confirms that spin-orbit interactions in the system are exceptionally strong \cite{Bergmann82b, Koga02}.

In order to proceed with fitting the data with the Hikami--Larkin--Nagaoka (HLN) theory \cite{Hikami}, we need to calculate the conductivity correction
\begin{equation}
\Delta \sigma (B)=[\sigma (B)-\sigma
(0)]-[\sigma_\mathrm{class}(B)-\sigma_\mathrm{class}(0)],
\end{equation}
where $\sigma$ is the longitudinal conductivity, obtained  from the inversion of the measured resistivity tensor, and $\sigma_\mathrm{class} $ is the classical longitudinal conductivity, obtained from the fitted $\rho_\mathrm{class}$. The obtained conductivity correction $\Delta \sigma (B)$ is plotted in Fig.\,4. The dots represent the measured data and the full lines are fits of the HLN-theory for the top-gate configurations $V_\mathrm{TG}=1$\,V, $k_\mathrm{F}l_\mathrm{m} =120$ (grey), and $V_\mathrm{TG}=0$, $k_\mathrm{F} l_\mathrm{m} =200$ (black).
Strictly speaking the HLN-theory is valid in the diffusive regime, where $B<B_\mathrm{tr}=\hbar/(2e{l_\mathrm{m}}^2$). In the case of the investigated sample we have $B_\mathrm{tr}<0.3$\,mT. The fitting interval shown in Fig.\,4 is taken to be slightly larger than this value in order to include a reasonable number of points. The data are fitted with the expression for strong SOI in the limit $B\ll B_\varphi$ \cite{Hikami, Altshuler81}
\begin{equation}
\Delta \sigma (B)= -\frac{e^2}{\pi h}
\displaystyle{\left(\frac{1}{2} \Psi
\left(\frac{1}{2}+\frac{B_{\varphi}}{B}\right)-\frac{1}{2}
\ln\frac{B_{\varphi}}{B}\right)},
 \label{c5_eq:Hikami1}
\end{equation}
where $\Psi(x)$ is the digamma function, $B_{\varphi}=\hbar/(4De\tau_{\varphi})$, $D$ is the diffusion constant and $\tau_{\varphi}$ is the phase-coherence time.
\begin{figure}
  \begin{center}
    \includegraphics{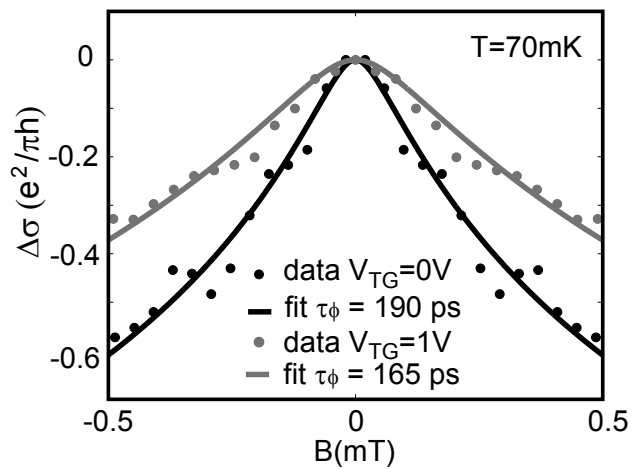}
    \caption{Fit of the anti-localization conductance peak with the HLN-theory [eq.\,(4)]---full lines are fitted, points are experimental data for the top-gate configurations $V_\mathrm{TG}=1$\,V, $k_\mathrm{F}l_\mathrm{m} =120$ (grey), and $V_\mathrm{TG}=0$, $k_\mathrm{F} l_\mathrm{m} =200$ (black). }
    \label{fig4NEW}
  \end{center}
\end{figure}
The only fitting parameter is $B_{\varphi}$. Satisfactory fitting is obtained (full lines in Fig. 4) for both top-gate configurations and the phase-coherence time of holes is extracted. In the configuration $V_\mathrm{TG}=1$\,V, $k_\mathrm{F} l_\mathrm{m} =120$ (gray points) we obtain $B_{\varphi}=5.1\times 10^{-5}$\,T, $\tau_{\varphi}=165$\,ps, and 
in the configuration $V_\mathrm{TG}=0$, $k_\mathrm{F} l_\mathrm{m} =200$ (black points) we obtain $B_{\varphi}=2.6\times 10^{-5}$\,T, $\tau_{\varphi}=190
$\,ps. The corresponding phase coherence length $l_{\varphi}$ of holes, calculated according to the diffusive regime expression $l_{\varphi}=\sqrt{D \tau_{\varphi}}$, are $1.8\,\mu$m and $2.5\,\mu$m, respectively. These values show that the phase-coherence length of holes decreases as the density in the sample is reduced. The values are compatible with those obtained from measurements of Aharonov--Bohm oscillations in p-type GaAs rings \cite{Grbic07}. They demonstrate that the fabrication of phase-coherent p-type GaAs nanostructures is accessible with present nanofabrication technologies. However, the values of $l_{\varphi}$ in hole systems are approximately one order of magnitude smaller than in electron systems with comparable densities and mobilities \cite{Ihn03,Hansen01}. Such a tendency was also observed in recent measurements of dephasing times of holes in open quantum dots \cite{Faniel06}. It suggests stronger charge dephasing in hole than in electron systems, presumably due to stronger carrier-carrier interactions \cite{Seelig01}.

Figure\,5(a) shows the temperature evolution of the resistivity around $B=0$ in the top-gate configuration $V_\mathrm{TG}=1$\,V, $k_\mathrm{F} l_\mathrm{m}=120$. The anti-localization dip depends strongly on temperature and disappears completely above 300\,mK, compatible with the temperature evolution of the Aharonov--Bohm oscillations in p-type GaAs rings \cite{Grbic07}. In addition, the resistance at $B=0$ exhibits metallic behavior with the zero-field resistivity increasing with temperature.

\begin{figure}
  \begin{center}
    \includegraphics{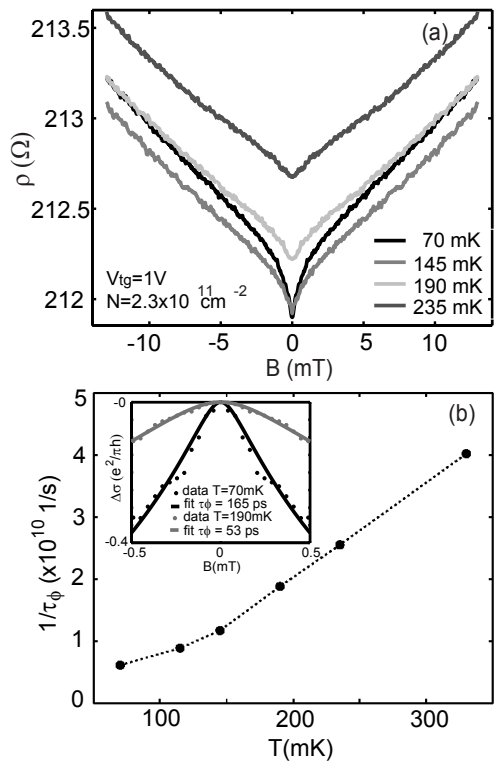}
    \caption{(a) Temperature dependence of the anti-localization resistivity minimum in the top-gate configuration $V_\mathrm{TG}=1$\,V, $k_\mathrm{F}l_\mathrm{m}=120$; (b) Temperature dependence of the inverse phase-coherence time of holes; Inset: fit of the anti-localization conductance peak with the HLN-theory [eq.\,(4)] for temperatures 70\,mK (black) and 190\,mK (gray)---full lines are fitted curves, points are experimental data.}
    \label{fig5NEW}
  \end{center}
\end{figure}
The fitting of the anti-localization peak in the conductance is performed for each measured temperature and the phase-coherence times are extracted. The inset of Fig.\,5(b) shows fits obtained for temperatures of 70\,mK and 190\,mK, from which the phase coherence times $\tau_{\varphi}=165$\,ps and $\tau_{\varphi}=53$\,ps, respectively,
are extracted. It can be seen in Fig.\,5(b) that the dephasing rate $\tau_{\varphi}^{-1}$ depends almost linearly on temperature.

Before we proceed with the evaluation of the spin-orbit scattering time $\tau_{SO}$ from weak anti-localization measurements we estimate $\tau_{SO}$ from SdH measurements. The estimated spin-orbit induced splitting of the heavy hole band at a density of $N=2.3\times 10^{11}$\,cm$^{-2}$ is $\Delta_\mathrm{SO}=0.47$\,meV [Fig.\,1(b)]. In semiconductor heterostructures with inversion asymmetry the dominant spin-orbit relaxation mechanism is the Dyakonov-Perel mechanism \cite{Dyakonov71}, which leads to the relation ${\tau_{SO}}^{-1}= {\Delta_\mathrm{SO}}^2\tau_\mathrm{tr}/4\hbar^2$ \cite{Dyakonov71}. Inserting the Drude transport scattering time $\tau_\mathrm{tr}=26$\,ps, we estimate $\tau_\mathrm{SO} \sim 0.3$\,ps. This shows that the SOI is so strong, that $\tau_\mathrm{SO} \ll \tau_\mathrm{tr}$. Therefore the SOI cannot be treated as a weak perturbation, which is the common assumption in theoretical calculations. An estimate of the characteristic field $B_\mathrm{SO}=\hbar/(4De\tau_{SO})$, at which 
the effects of SOI become suppressed and the weak anti-localization positive magnetoresistance turns into a weak localization negative magnetoresistance \cite{Knap96} gives $B_\mathrm{SO}\sim 30$\,mT, which is far beyond the transport field $B_\mathrm{tr} \sim 0.3$\,mT up to which diffusive theories of weak anti-localization are applicable. However, the value $B_\mathrm{SO} \sim 30$\,mT provides a qualitative
understanding of the fact that we observe only a weak anti-localization dip without a weak localization peak in the measured magnetoresistivity.

We show the results of fitting the data in a wide magnetic field range with the HLN-theory using the expression \cite{Knap96, Iordanskii94, Hikami}
\begin{eqnarray}
\Delta \sigma (B)= -\frac{e^2}{\pi h} \left[ \frac{1}{2} \Psi
(\frac{1}{2}+\frac{B_{\varphi}}{B})-\frac{1}{2}\ln
\frac{B_{\varphi}}{B}\right. \nonumber\\
-\Psi (\frac{1}{2}+
\frac{B_{\varphi}+B_{SO}}{B}) + \ln \frac{B_{\varphi}+B_{SO}}
{B} \nonumber \\
\left.- \frac{1}{2} \Psi (\frac{1}{2}+\frac{B_{\varphi}+2B_{SO}}{B}) +\frac{1}{2}\ln
\frac{B_{\varphi}+2B_{SO}}{B}\right]
 \label{c5_eq:Hikami2}
\end{eqnarray}

It should be mentioned that the HLN-theory was originally developed for metallic samples where the Elliot SO skew-scattering mechanism is present. For this mechanism the spin-splitting energy is proportional to $k^3$. However, in most semiconductor heterostructures the Dyakonov-Perel spin relaxation is dominant \cite{Dyakonov71}. The theory by Iordanskii--Lyanda-Geller--Pikus (ILP) describes the weak anti-localization correction for this type of spin relaxation, and it involves both linear and cubic in $k$ spin-splitting terms \cite{Iordanskii94}. If the linear contribution is negligible and cubic spin-splitting is dominant, the ILP-theory gives the same result as the HLN-theory in eq.\,(5) \cite{Iordanskii94, Knap96}. Since the spin-orbit induced splitting of the heavy hole GaAs band is proportional to $k^3$ \cite{Gerchikov92, Winkler03}, it is appropriate to use the HLN eq.\,(5) for fitting the weak anti-localization in hole systems. Equation\,(5) contains two fitting parameters, namely $B_{\varphi}$ and $B_\mathrm{SO}$.
\begin{figure}
  \begin{center}
    \includegraphics{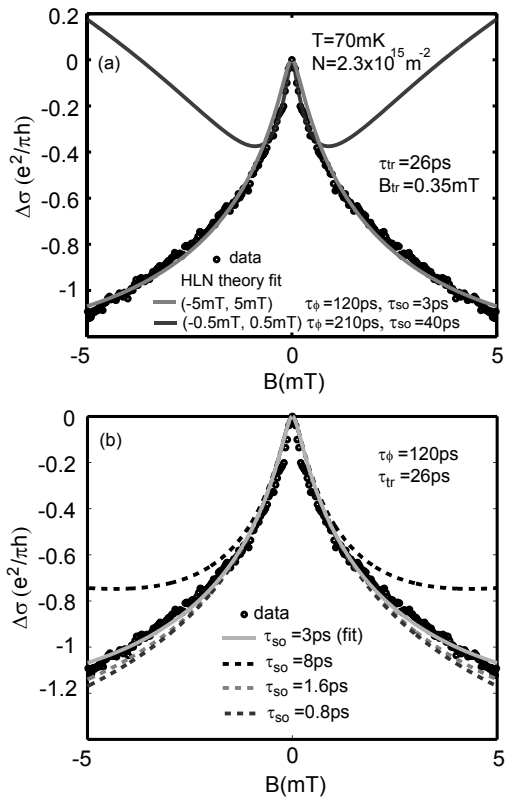}
    \caption{(a) Fits of the weak anti-localization conductance peak with the HLN-theory including SOI [eq.\,(5)] in the range $|B|<0.5$\,mT (dark gray line) and in the range $|B|<5$\,mT (light gray line). (b) Sensitivity of the fits in the range $|B|<5$\,mT to the change of $\tau_\mathrm{SO}$ (the full grey line represents the fit obtained in (a), while dashed lines correspond to different values of $\tau_\mathrm{SO}$ as quoted in the figure).}
    \label{fig6NEW}
  \end{center}
\end{figure}

The sharpness of the weak anti-localization conductance peak is determined by $\tau_{\varphi}$, whereas the tail of the peak depends on $\tau_\mathrm{SO}$ \cite{Altshuler81, Knap96}. Therefore we explore in Fig.\,6(a) how the fit of the data with eq.\,(5) depends on the magnetic field range. Fitting in the narrow range $|B|<0.5$\,mT [dark gray line in Fig.\,6(a)] reproduces the low field behavior quite well up to $B\sim B_\mathrm{tr}$, but above that field the fit does not match the experimental data. The obtained fitting parameters are in this case $\tau_{\varphi}=210$\,ps and $\tau_\mathrm{SO}=40$\,ps. On the other hand the fit of the data in the larger range $|B|<5$\,mT [light grey line in Fig.\,6(a)] matches the tails of the peak better, but does not fit the low-field data below $B_\mathrm{tr}$ satisfactorily. The fit parameters in this second case are $\tau_{\varphi}=120$\,ps and $\tau_\mathrm{SO}=3$\,ps. While the obtained values for $\tau_{\varphi}$ differ by less than a factor of 2, and are comparable with the value obtained by fitting with  eq.\,(4) which neglects the contribution of $\tau_\mathrm{SO}$, the obtained values for $\tau_\mathrm{SO}$ differ by more than an order of magnitude. Although the fit in the range $|B|<0.5$\,mT, i.e., below $B_\mathrm{tr}$, is theoretically more justified, it is clear that it underestimates the SO strength, because it gives an up-turn from weak anti-localization to weak localization which is not observed in the measured data. Also the value $\tau_\mathrm{SO}=40$\,ps is significantly larger than that estimated from the beating of the SdH oscillations $\tau_\mathrm{SO}=0.3$\,ps.
Fitting in the range $|B|<5$\,mT gives better agreement between the extracted $\tau_\mathrm{SO}=3$\,ps and the value obtained from SdH oscillations. In Fig.\,6(b) we investigate the influence of changing $\tau_{SO}$ at fixed $\tau_{\varphi} =120$\,ps on the fitted curves and find that the fitting procedure becomes less sensitive for $\tau_\mathrm{SO}< 3$\,ps. Therefore, rather than giving the exact value, this fit sets the upper limit on $\tau_\mathrm{SO}$.

It should be mentioned that even admitting anisotropic spin relaxation and using the theory of Ref.\,\onlinecite{Averkiev98} with three instead of two fitting parameters did not give better fits to the data. Also, curves simulated using the theory developed for the ballistic regime \cite{golub05} could not satisfactorily match the data in the entire investigated magnetic field range.

Due to the exceptionally strong SOI effects and the high mobility of holes in our p-type GaAs sample, it is in the regime where $\tau_\mathrm{SO} \ll \tau_\mathrm{tr} \ll \tau_{\varphi}$ ($\tau_\mathrm{SO}\sim 0.3$\,ps, $\tau_\mathrm{tr}=26$\,ps, $\tau_{\varphi}=190$\,ps). In this regime SOI cannot be treated perturbatively, as it is the case in the more commonly studied regime $\tau_\mathrm{tr} \ll \tau_\mathrm{SO} \ll \tau_{\varphi}$, where a small and sharp weak anti-localization resistance minimum is superimposed on a wider weak localization resistance peak. This might explain the difficulties in fitting our weak anti-localization data in a larger magnetic field range with present theoretical models. Similar difficulties in fitting weak anti-localization data were observed for an InGaAs/InP quantum well with strong SOI \cite{studenikin}.

It is also possible that the difficulties in fitting the weak anti-localization data arise from the fact that the low-field magnetoresistance contains some other contribution, in addition to the weak anti-localization, presumably due to carrier--carrier Coulomb interactions \cite{Altshuler81}. These interaction-corrections might be particularly strong in p-type
GaAs systems due to the effective mass of holes which is significantly larger than in n-GaAs systems.

\section{Conclusions}

In conclusion, we have performed a detailed analysis of the low-field magnetoresistance in a carbon doped p-type GaAs heterostructure. The presence of exceptionally strong spin--orbit interactions in the structure is demonstrated by the simultaneous observation of a beating of SdH oscillations, a classical positive magnetoresistance and a weak anti-localization correction in the magnetoresistance. A spin--orbit induced heavy-hole subband splitting of around 30\% of the Fermi energy is deduced from the beating of SdH oscillations. The classical positive magnetoresistance, originating from the presence of the two spin-split subbands, has been fitted with a two-band model up to the fields where SdH oscillations are not yet developed, allowing to estimate the inter- and intra-subband scattering rates. In a very narrow  magnetic field range around $B=0$, a weak
anti-localization resistivity minimum is observed. The fact that this minimum is not superimposed on a  wider weak localization peak confirms that the sample is in the regime where $\tau_\mathrm{SO} \ll \tau_\mathrm{tr} \ll \tau_{\varphi}$, i.e., where spin--orbit interactions are very strong and can not be treated perturbatively in calculations of quantum corrections of the magnetoresistance. From weak anti-localization measurements the phase-coherence time of the holes is determined to be around 190\,ps at $T=70$\,mK. The temperature dependence reveals that the weak anti-localization resistance minimum persists up to 300\,mK and that $\tau_{\varphi}^{-1}$ depends
on temperature in an almost linear fashion. The extracted phase coherence length of holes of around 2.5\,$\mu$m at $T=70$\,mK shows that the fabrication of phase-coherent p-type GaAs nanodevices is possible using present nanofabrication technologies.

We thank L. Golub and M. Glazov for stimulating discussions.
Financial support from the Swiss National Science Foundation is
gratefully acknowledged.

\end{document}